\documentclass[aps,prl,twocolumn,groupedaddress,floatfix]{revtex4}
\usepackage{graphicx,color}
\usepackage{amsmath}
\usepackage{amssymb}
\usepackage{bm}
\begin{document}
\title{Two types of topological transitions in finite Majorana wires}
\author{Dmitry I. Pikulin}

\affiliation{Instituut-Lorentz, Universiteit Leiden, P.O. Box 9506, 2300 RA Leiden, The Netherlands}

\author{Yuli V. Nazarov}

\affiliation{Kavli Institute of Nanoscience, Delft University of Technology, Lorentzweg 1,
2628 CJ Delft, The Netherlands}

\begin{abstract}
Motivated by the recent advances in studying Majorana states in nanowires under conditions of superconducting proximity effect, we address the correspondence between the common topological transition in infinite system and the topological transition of other type that manifests itself in the positions of the poles of the scattering matrices. We establish a universal dependence of the pole positions in the vicinity of the common transition on the parameter controlling the transition, and discuss the manifestations of the pole transitions in the differential conductance.   
\end{abstract}
\maketitle


Majorana bound states have been predicted to exist in various condensed matter setups:  5/2 FQHE state \cite{Moore}, in vortices found in $p+ip$ superconductors \cite{Ivanov} and in specific models of 1d superconductors\cite{Kitaev1}. The importance of the Majorana states for quantum computation \cite{Kitaev2} has brought them to the focus of the condensed matter research\cite{Beenakker}.
Next step were the suggestions to realize the Majorana states in more experimentally feasible setups, those include topological insulators \cite{Fu, Akhmerov} and semiconductor nanostructures with big spin-orbit interaction brought in proximity to s-wave superconductors.  Two- \cite{Sato, Sau} and one-dimensional \cite{Lutchyn, Oreg} nanostructures have been considered. 

The observation of Majorana bound states in 1d  nanowires has been reported by several groups by measuring zero-bias conductance peak \cite{Mourik, Deng, Das} and  4$\pi$ Josephson effect \cite{Rokhinson}. The signature of Majorana's is their emergence upon a topological transition \cite{Volovik} separating the regions of parameter space with and without zero-energy excitations.  In all cases the experiments have been performed with finite and rather short wires. This brings about the question: how a topological transition taking place in infinite system is manifested in properties of a finite wire. 

Strictly speaking, this common topological transition is absent in a finite system where excitation energies continuously depend on the control parameter of the transition and are never precisely zero \cite{Us, Kitaev1, Lutchyn}. This may be shown in several ways. In \cite{Us} we gave the most general formulation in terms of  the topology of the energy-dependent scattering matrix characterising a finite nanostructure.
Same study revealed a topological transition of other kind that takes place in finite systems and manifests itself in the properties of the poles of the scattering matrix. The topological number in this case is the number of poles at purely imaginary energy, and the topological transition is the change of this integer even number upon the continuous variation of the control parameter. 

In the present work we link these two topological transitions of different types: bulk one and finite system one. We show that in general case the common topological transition is accompanied by the pole topological transition\cite{opposite}. The points of the transitions differ at the scale inversely proportional to the wire length. We implement the generic model of the Majorana wire that is always valid in the vicinity of the transition point and obtain the universal dependence of the pole positions on the control parameter and a single parameter characterizing the coupling of the wire to a normal metal lead.  We discuss how the same correspondence occurs for more specific models and how the universal picture is manifested in a transport measurement.

The "standard" model describing a Majorana wire encompasses a single-band spectrum that includes spin-orbit interaction, proximity effect from the bulk superconductor and spin magnetic field\cite{Lutchyn, Oreg}. Let us derive a phenomenological effective model valid near the common topological transition point. We can start with a multi-mode wire where the spectrum at each $k$ is described by a general Hamiltonian matrix $\hat{H}(k)$ in the space of the modes and Nambu index. The general symmetry of BdG equations \cite{BdGsymmetry} requires $\hat{H}(k) = -\hat{H}^T(-k)$ in a certain (Majorana) basis. The common topological transition takes place when an eigenvalue of $\hat{H}(k=0)$ passes $0$ indicating a closing of the superconducting proximity gap in the wire.

Owing to BdG symmetry, the zero eigenvalue is doubly degenerate. Thus we concentrate at the two modes corresponding to the eigenvalue.
Near the transition point the general form for this Hamiltonian in Majorana basis reads
$H(k=0) = a \sigma_y$, where $\sigma$'s here and below are usual Pauli matrices. The phenomenological parameter $a$
controls the proximity to the transition and is a function of physical control parameters like magnetic field or chemical potential, $a=0$ in the transition point. Expanding near $k=0$  
 and taking into account the BdG symmetry, we find two possible terms $\propto k \sigma_x$ and $\propto k \sigma_z$. The combination of the two can be brought to $\propto k\sigma_z$  by a rotation of the pseudospin about $y$ axis. This brings us to the generic Hamiltonian we will use in further consideration:
\begin{equation}
H = v k \sigma_z + a \sigma_y.
\label{eq:H}
\end{equation}
It has been first introduced in \cite{Akhmerov1}.

\begin{figure}
\includegraphics[width=0.9\linewidth]{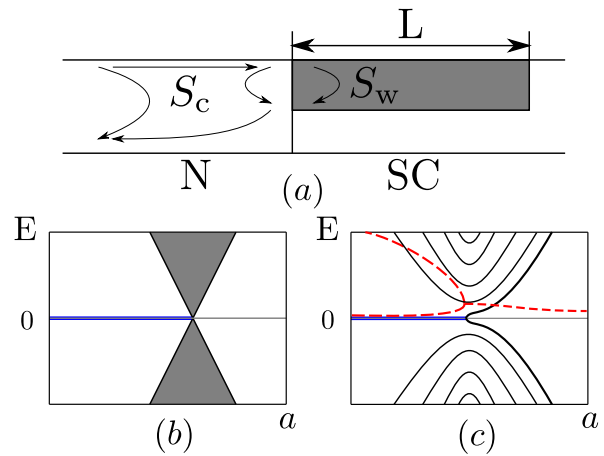}
\caption{(a) The setup: the Majorana wire of the length $L$ at the top of a superconductor is connected to a normal-metal lead. The total scattering matrix at low energy incorporates that of the contact  ($S_{{\rm c}}$) and energy-dependent scattering matrix describing propagation in the  wire, $S_{\rm w}$. (b) Sketch of the (continuous) spectrum in the limit of infinite $L$: a Majorana level emerges upon the {\it common} topological transition. (c) The common topological transition becomes a crossover for finite $L$. The quantized energy levels (real parts of the pole energy positions)
are sketched versus the control parameter $a$. The lowest level reaches $0$ at the point of the {\it pole} topological transition. Dashed lines give the imaginary parts of the pole positions for the lowest level.} 
\label{fig1}
\end{figure}

Let us turn to a finite wire The boundary conditions at the wire ends must be consistent with the current conservation. The operator of current reads $\hat I = v \sigma_z$ so the conservation implies that the wavefunction $\Psi = \left\{\psi_1, \psi_2\right\}^T$ has to satisfy: 
\begin{equation}
|\psi_1|^2 = |\psi_2|^2,
\end{equation}
Near zero energy the wavefunction is real, and we are left with binary choice $\psi_1 = \pm \psi_2$. We fix the signs to $+$ at the right end of the system and $-$ at the left one. In this case, in the limit of infinite wire length the Majorana states are formed at $a <0$ while the phase at $a >0$ is topologically trivial (Fig. \ref{fig1}b).

Let us now contact the left end of the wire with a normal metal lead and describe the situation in therms of the scattering matrix from/to normal lead modes. Scattering matrices are very useful objects to study the properties of the superconducting junctions\cite{Shelankov}. They incorporate relevant details of the setup in few parameters and allow to compute different properties of the junction, like conductivity. They also allow for the topological classification of the junction in a concise way \cite{Us}. 

There are two interesting modes in the wire propagating in opposite directions. The scattering matrix of the contact $S_{\rm c}$ is in the basis of the incoming waves in the lead and the single mode of the wire and is thus of size $M+1 \times M+1$, $M$ is the number of modes in the normal lead. We separate it into blocks as:
\begin{equation}
S_{\rm c} = \left(
\begin{array}{cc}
\check{r}_{11} & \check{r}_{21}\\
\check{r}_{12} & - r
\end{array}
\right).
\end{equation}
Here $\check{r}_{11}$ is $M\times M$ matrix of the (Andreev)reflection to the leads that also incorporates the scattering in all other wire modes, $\check{r}_{21, \, 12}$ are scattering amplitudes from/to the wire and $r$ is a number, which gives the reflection amplitude in the wire ($r=1$ corresponding to the wire isolation)\cite{rsign}. By virtue of BdG condition $r$ is real at zero energy. Since the interesting energy dependence comes from the wire, we can neglect the energy dependence of $S_{\rm c}$.  

To get the full $M\times M$ scattering matrix in the space of normal lead modes, we need to combine the $S_{\rm c}$ with the scattering amplitude $S_{\rm w}$ that describes the propagation along the wire, reflection from the right and the propagation back to the left end. This amplitude is easy to find from the Hamiltonian (1) and reads

\begin{equation}
S_{\rm w} = e^{i \chi} = \frac{\coth(\sqrt{a^2 - \epsilon^2} L/v ) + \frac{a+i\epsilon}{\sqrt{a^2 - \epsilon^2}}}{\coth(\sqrt{a^2 - \epsilon^2} L/v ) + \frac{a-i\epsilon}{\sqrt{a^2 - \epsilon^2}}}.
\label{eq:poles0}
\end{equation}
$L$ being the wire length. 
The whole peculiarity of the limit of infinite wire may be seen from this formula. If we formally set $L\to \infty$ we get  $S_{\rm w} = {\rm sign}(a)$. The matrix is thus energy-independent and topologically trivial (nontrivial) for $a<0$ ($a>0$). 

The full scattering matrix thus reads 
\begin{equation}
\label{eq:S}
S = \check r_{11} + \check r_{21} e^{i\chi} \frac{\,}{1 + r e^{i\chi} } \check r_{12}.
\end{equation}
We concentrate on {\it poles} of this matrix those are solutions of  
\begin{equation}
{\sqrt{a^2 - \epsilon^2}} \coth(\sqrt{a^2 - \epsilon^2} L/v ) + {a-i\epsilon \frac{1-r}{1+r}} = 0.
\end{equation}

\begin{figure}
\includegraphics[width=0.9\linewidth]{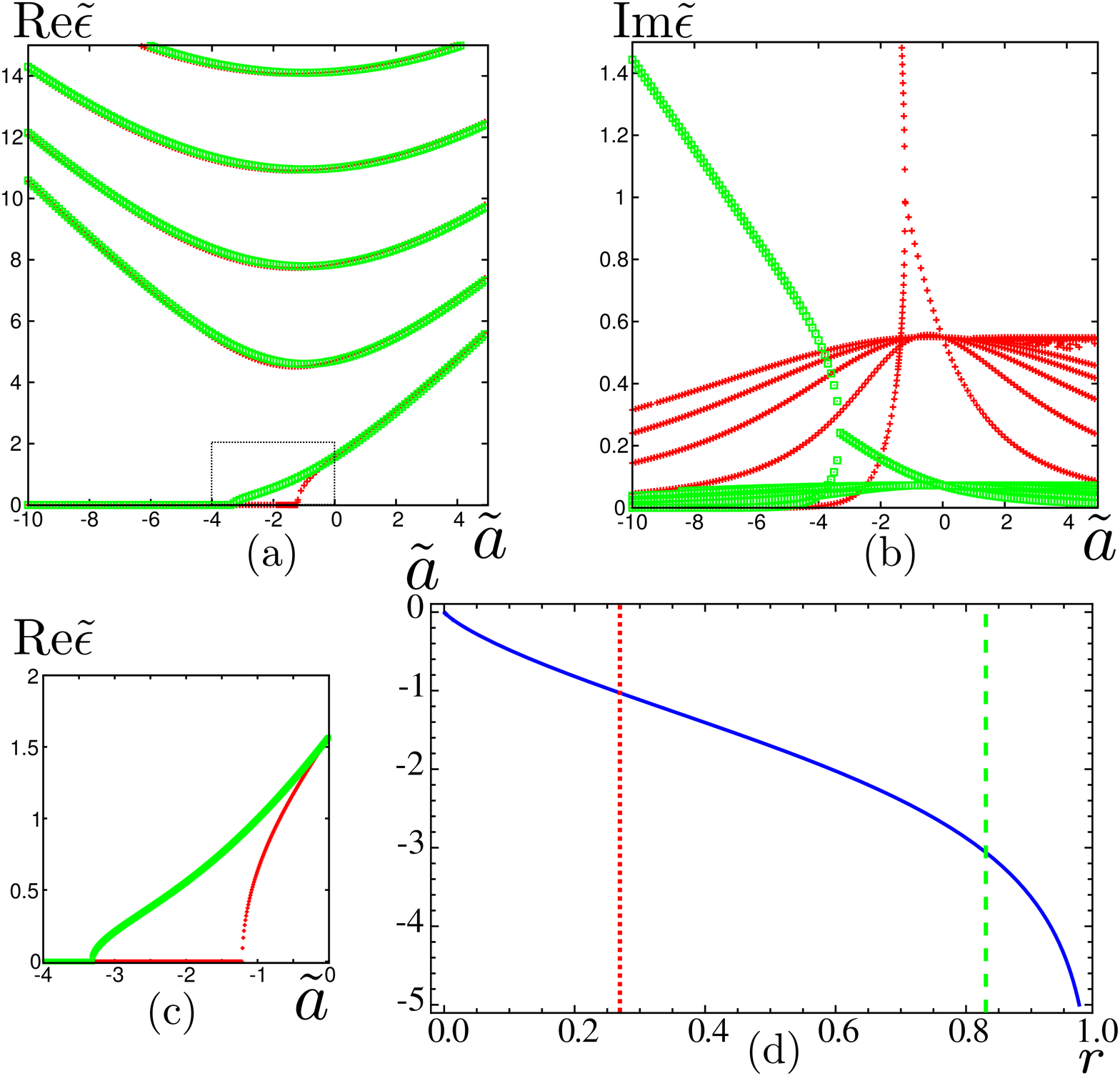}
\caption{The real (a) and and imaginary (b)  parts of the pole energy positions versus the control parameter $\tilde{a}$ at two values of the reflection amplitude $r$: $r=0.86$ (green dots, almost isolated)and $r=0.34$ (red crosses, almost transparent). The part of (a) within the rectangular is replotted in (c).
The pole topological transition occurs at $a_0 = -1.2$ for $r=0.86$ and $a_0 = -3.3$ for $r=0.34$. (d)
 The dependence of the transition point $a_0$ on $r$.} 
\label{fig2}
\end{figure}

At finite length $L$ the common topological transition becomes a crossover taking place in an interval of $a$ of the order of effective level spacing in the wire $v/L$ and at the corresponding energy scale. We aim to describe this universal crossover. To this end we rescale $a,\epsilon$ to dimensionless units $\tilde a = a \frac{v}{L}$, $\tilde \epsilon = \epsilon \frac{v}{L}$. The equation becomes
\begin{equation}
\sqrt{\tilde a^2 - \tilde \epsilon^2} \coth(\sqrt{\tilde a^2 - \tilde \epsilon^2} ) + 
\tilde a-i\tilde \epsilon \frac{1-r}{1+r}= 0.
\label{eq:all_energies}
\end{equation} 
Numerical solutions for pole positions are shown in Fig. \ref{fig2}a,b for two values of $r$ as functions of the control parameter $\tilde a$. We see a sharp feature in the crossover region: the pole topological transition. At this point, the real part of the energy of the lowest pole becomes strictly zero. This occurs at finite negative values of $\tilde a$.  The higher the transmission through the $S_{c}$, the closer to $0$ is the transition point. 
This dependence is presented in Fig.\ref{fig2}c.
In the limit of low transmissions, the pole transition takes place at $|\tilde{a}| \simeq \ln(1-r)$  where the exponentially small splitting of Majorana states matches small decay rate of the left-end state to the normal metal. The real parts of energies of all other poles follow the hyperbola-like curves indicating formation of discrete energy levels in the wire above the gap edge $|\tilde{a}|$. The same transition is seen in imaginary parts of energy positions as a splitting of the curve corresponding to the lowest pole. The  upper (lower) parts of the split curve give the decay rates of the left(right) end Majorana state. The decay rate for the Majorana "buried" at the right end falls off exponentially with increasing $|\tilde a|$: $\tilde{\epsilon} \approx 2 i \tilde{a} \exp( - 2 |\tilde{a}|) \frac{1+r}{1-r}$.

Let us address the dependence of the pole positions on $a$ near the pole transition point. For this, we expand \eqref{eq:all_energies} near the transition point $\tilde a=a_0$, $\tilde\epsilon=i \epsilon_0$ to obtain the relation between the deviations $\delta \epsilon$,$\delta a$ from the transition point in the lowest non-vanishing order:
\begin{widetext}
\begin{equation}
\delta a = C \delta \epsilon^2;\;
C=\frac{-a_0^3 (1+2 \varepsilon_0  \mu )-2 \varepsilon_0 ^5 \mu  \left(-1+\mu ^2\right)+2 a_0 \varepsilon_0 ^2 \left(1+\varepsilon_0 ^2 \left(1-3 \mu ^2\right)\right)+a_0^2 \varepsilon_0  \left(-3 \mu -\varepsilon_0  \left(-1+4 \varepsilon_0  \mu +\mu ^2\right)\right)}{2 \varepsilon_0  \left(a_0^2+\varepsilon_0 ^2\right) \left(a_0 (1+2 a_0) \mu +\varepsilon_0  \left(-1+a_0 \left(-1+\mu ^2\right)\right)\right)} \simeq 1.
\end{equation}
\end{widetext}
Here $\mu = \frac{1-r}{1+r}$. 
This gives square root splitting of either real parts of the energy positions $\delta \epsilon = \pm \sqrt{\delta a /C}$ at $\delta a >0$ or imaginary ones,  $\delta \epsilon = \pm i \sqrt{|\delta a|/C}$ at $\delta a <0$.
This square root dependence of $\delta \epsilon$ on $\delta a$ is in full agreement with Fig. \ref{fig2}a,b.

\begin{figure}
\includegraphics[width=0.9\linewidth]{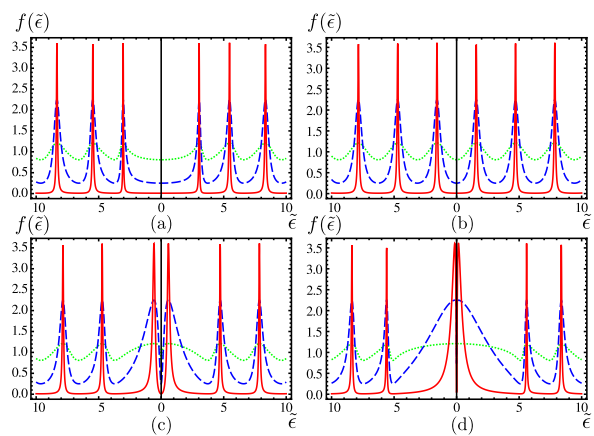}
\caption{The universal bias-dependent contribution to differential Andreev conductance of the contact versus energy/voltage. Solid, dashed, dotted curves correspond to $r=0.9,0.5,0.1$ respectively. (a)  $\tilde{a}=2$, long before the transition. (b) $\tilde{a}=0$(c) $\tilde{a}= -2$ in the crossover interval. (d) $\tilde{a}= -4$ long after the transition. The Majorana at the far end of the wire is manifested as a narrow dip at zero bias.} 
\label{fig3}
\end{figure} 

The experimentally observable quantity is the differential conductance of the contact, $G(\epsilon = eV)$, $V$ being the voltage drop at the contact. 
In terms of the scattering matrix, the conductance reads 
$G = \frac{e^2}{2\pi\hbar} {\rm Tr} ( 1 - \sigma_y S \sigma_y S^\dag)$. Substituting $S$ in the form of (\ref{eq:S}), we obtain a universal energy dependence of the conductance in the crossover region,
\begin{eqnarray}
G(\epsilon) = G_0 + G_1 f(a,\epsilon); \\ f(a, \epsilon) = \frac{(1-r^2)^2}{1 + r^2 + 2 r \cos \chi (a,\epsilon)}.
\end{eqnarray}
The dependence is governed by the universal function $f(a,\epsilon)$ ($0<f<4$) while the non-universal coefficients $G_0, G_1$ depend on the details of the $S_{{\rm c}}$,
\begin{eqnarray}
G_0 = \frac{e^2}{2\pi\hbar}{\rm Tr} \left(1 - ( \check{r}_{11} + r^{-1} \check{r}_{21} \check{r}_{12} ) \sigma_y\check{r}_{11}^T\sigma_y\right),\nonumber\\
G_1 = - \frac{e^2}{2\pi\hbar}\frac{{\rm Tr} \left(\sigma_y \check{r}_{21} \check{r}_{12} \sigma_y( \check{r}_{12}^T \check{r}_{12}^T + (r - r^{-1}) \check{r}_{11}^T)\right)}{(1-r^2)^2}.
\end{eqnarray}
The coefficient $G_1 \simeq e^2/\hbar$ and can be of {\it any} sign while $G_0$ can be much bigger than $e^2/\hbar$.
The function $f$ (Fig. \ref{fig3}) at any $r$ gives a sequence of peaks associated with the poles of the scattering matrix. The peaks are narrow in the isolation limit $r \to 1$. Before the transition, the peaks are far from zero energy. Upon the crossover, the peaks come close to zero
and almost merge near the transition point. However, they never merge to a single peak: the Majorana state at far end of the wire is manifested in the conductance as a dip that becomes increasingly narrow upon increasing $-\tilde{a}$. 

Since the poles always have a finite imaginary part, and the conductance is defined at real energy, there is no singularity in $f(\epsilon)$ at the point of the pole topological transition. However, this singularity can be in principle identified from the experimental data by numerical analytical continuation to complex energy plane. The pole topological transition in principle does not require the common topological transition and can occur, for instance, in strongly disordeded wires \cite{disorder}. 
The signatures of pole transitions have been investigated for the disordered wire model \cite{Pikulin} in the parameter region where no common transition can guarantee the presence of Majorana mode.

Another setup proposed \cite{Akhmerov1} to reveal the signatures of Majorana fermions encompasses normal leads at both ends of a finite nanowire. Also in this case the common topological transition is accompanied by a pole transition and proceeds in a similar way. The qualitative difference is that far below the transition both Majorana states retain a finite width and each of the two associated poles is manifested only in the scattering from the corresponding end of the wire.
In the model under consideration, the Majorana splitting retains the same sign. More detailed models, e.g.\cite{Lutchyn}, predict spectacular oscillations of the splitting \cite{Klinovaja}. We stress that in the limit of the long wires $Lk_F \gg 1$ such oscillations can only start far from the common topological transition, that is, at the values of the control parameter that are parametrically bigger than $v/L$.

To conclude, we have formulated and studied a universal model that describes the crossover in the vicinity of the common topological transition for finite clean Majorana wires. Importantly, we have shown that the sharp pole topological transition takes place in the crossover interval of the control parameter and computed the dependence of the pole positions on the contol parameter in this interval. We have also found a universal shape of differential conductance for this model, this enables its starightforward experimental verification. 

We stress the universal character of our conclusions, in particular, the predictions for the conductance: those should hold in any sufficiently long wire with small disorder in the vicinity of the topological transition. Some features of our results have been seen in Ref. \cite{Mourik}: the authors have observed a narrow zero-bias peak on the background of a wider dip as seen in Fig. \ref{fig3}d (assuming $G_1$ is negative).
From the other hand, no regular patten of peaks moving to zero upon changing the control parameter has been observed so far. More experimental data, in particular, for longer wires are required to clarify the discrepancy that can be due to sufficiently strong disorder or finite temperature effects.

This research was supported by the Dutch Science Foundation NWO/FOM. The authors are grateful to C. W. J. Beenakker, P. A. Ioselevich, and A. Cottet for useful discussions.


\begin{thebibliography}{100}
\bibitem{Moore}
G. Moore and N. Read, Nucl. Phys. B \textbf{360}, 362 (1991).

\bibitem{Ivanov}
D. A. Ivanov, Phys. Rev. Lett. \textbf{86}, 268 (2001).

\bibitem{Kitaev1}
A. Y. Kitaev, Phys.-Usp. \textbf{44}, 131 (2001).

\bibitem{Kitaev2}
A. Y. Kitaev, Ann. Phys. \textbf{303}, 2 (2003).

\bibitem{Beenakker}
See C. W. J. Beenakker, arXiv:1112.1950 for a review.

\bibitem{Fu}
Liang Fu and C. L. Kane, Phys. Rev. Lett. \textbf{100}, 096407 (2008).

\bibitem{Akhmerov}
A. R. Akhmerov, Johan Nilsson, and C. W. J. Beenakker, Phys. Rev. Lett. \textbf{102}, 216404 (2009).

\bibitem{Sato}
M. Sato, Y. Takahashi, S. Fujimoto, Phys. Rev. Lett. \textbf{103}, 020401 (2009).

\bibitem{Sau}
J. D. Sau, R. M. Lutchyn, S. Tewari, and S. Das Sarma, Phys. Rev. Lett. \textbf{104}, 040502 (2010).

\bibitem{Lutchyn}
R. M. Lutchyn, J. D. Sau, S. Das Sarma, Phys. Rev. Lett. \textbf{105}, 077001 (2010).

\bibitem{Oreg}
Y. Oreg, G. Refael, and F. von Oppen,  Phys. Rev. Lett. {\bf 105}, 177002 (2010).

\bibitem{Mourik}
V. Mourik, K. Zuo, S. M. Frolov, S. R. Plissard, E. P. A. M. Bakkers, and L. P. Kouwenhoven, Science \textbf{336}, 1003 (2012).

\bibitem{Deng}
M. T. Deng, C. L. Yu, G. Y. Huang, M. Larsson, P. Caroff, and H. Q. Xu, unpublished, arXiv:1204.4130.

\bibitem{Das}
A. Das, Y. Ronen, Y. Most, Y. Oreg, M. Heiblum, and H. Shtrikman, arXiv:1205.7073.

\bibitem{Rokhinson}
L. P. Rokhinson, X. Liu, J. K. Furdyna, Nature Physics doi:10.1038/nphys2429 (2012).

\bibitem{Volovik}
G. E. Volovik, \textit{The Universe in a Helium Droplet}, Oxford University Press (2003).

\bibitem{Us}
D. I. Pikulin and Yu. V. Nazarov, JETP Letters, \textbf{94}, 9, 693-697 (2011).

\bibitem{opposite}
On the contrary, the pole topological transition is not always accompanied by the bulk one. In \cite{Pikulin} an example of the pole topological transition deep inside trivial region of the bulk parameters was demonstrated.

\bibitem{Pikulin}
D. I. Pikulin, J. P. Dahlhaus, M. Wimmer, H. Schomerus, C. W. J. Beenakker, arXiv:1206.6687.

\bibitem{BdGsymmetry}
P.-G. de Gennes, \textit{Superconductivity of Metals and Alloys}, Addison-Wesley, Reading, MA (1986).

\bibitem{Akhmerov1}
A. R. Akhmerov \textit{et al.}, Phys.Rev.Lett. \textbf{106}, 057001 (2011).

\bibitem{Shelankov}
A. L. Shelankov, Zh. Eksp. Teor. Fiz., Pis’ma, {\bf 32}, 2, 122-125 (1980); G. E. Blonder, M. Tinkham, and T. M. Klapwijk, Phys. Rev. B {\bf 25}, 4515 (1982).

\bibitem{rsign}
We should also require $0\leq r < 1$, since  $r$ passing 0 would have indicated another topological transition, in this case  somewhere outside the wire \cite{Akhmerov1} and not related to it. 

\bibitem{disorder}
P. W. Brouwer, M. Duckheim, A. Romito, and F. von Oppen, Phys. Rev. B {\bf 84}, 144526 (2011).

\bibitem{Klinovaja}
J. Klinovaja and D. Loss, Phys. Rev. B \textbf{86}, 085408 (2012).

\end{thebibliography}
\end{document}